\documentstyle[12pt,aasms4]{article}
\input psfig
 
\begin{document}
 
\long\def\Ignore#1{\relax}
\def\spose#1{\hbox to 0pt{#1\hss}}
\def\ltsim{\mathrel{\spose{\lower.5ex\hbox{$\mathchar"218$}}
     \raise.4ex\hbox{$\mathchar"13C$}}}
\def\gtsim{\mathrel{\spose{\lower.5ex \hbox{$\mathchar"218$}}
     \raise.4ex\hbox{$\mathchar"13E$}}}
 
\title{On The Formation of Disk Galaxies and Massive Central Objects}
 
\author{J.\ A.\ Sellwood\footnotemark[1] and E.\ M.\ Moore}
\affil{Department of Physics and Astronomy, Rutgers, The State University of
New Jersey, 136 Frelinghuysen Road, Piscataway, NJ 08854-8019}
\affil{sellwood@physics.rutgers.edu}
 
\addtocounter{footnote}{1}
\footnotetext{\it Isaac Newton Institute for Mathematical Sciences, Cambridge
University, 20 Clarkson Road, Cambridge CB3 0EH, UK}
%\addtocounter{footnote}{2}
%\footnotetext{\it NASA }
 
\begin{abstract} We propose that massive central objects form in the centers
of the bars which must develop in young high-surface density galactic disks.
Large-scale dynamics shuts off the growth of the central mass before it
reaches $\sim 2$\% of the disk mass at the time, but this mass is sufficient
to weaken the bar substantially.  Subsequent evolution of the galaxy can
either complete the destruction of the bar or cause it to recover, depending
upon the angular momentum distribution of later infalling material.  We
produce massive, fully self-gravitating disks having roughly flat rotation
curves which are quite stable.  If at least part of the central masses we
require constitute the engines of QSOs, then our picture naturally accounts
for their redshift dependence, since the fuel supply is shut off by the
development of an inner Lindblad resonance.  A prediction is that massive
objects should not be found in halo dominated galaxies, such as
low-luminosity, or low-surface brightness galaxies.
 
\end{abstract}
 
\section{Introduction}
It is widely believed that galaxy disks form as gas cools and settles into
rotational balance in massive dark matter halos (White \& Rees 1978; Fall \&
Efstathiou 1980; Gunn 1982; Ryden \& Gunn 1987; van der Kruit 1987).  This
broad picture has been further developed by Dalcanton, Spergel \& Summers
(1997) and Mo, Mao \& White (1998) who, however, continue to employ Mestel's
(1963) simplifying assumption that infalling matter conserves its detailed
angular momentum distribution.  Such a treatment is inadequate once the disk
becomes massive enough to be self-gravitating because angular momentum will
be redistributed by non-axisymmetric instabilities.  While the consequences
of such instabilities have not been fully worked out, Dalcanton et al.\
plausibly suggested that they could be responsible for the formation of
bulges.  On the other hand, Mo et al.\ conclude that high mass disks are
unphysical on the obscure grounds that ``only dynamically stable systems can
correspond to real galaxy disks.''
 
If the disk in a high-surface brightness (HSB) galaxy is in fact the dominant
mass component in the inner parts, for which there is considerable evidence
(see \S2), then it must have become self-gravitating at an early stage.  It
is therefore an issue of some urgency to determine what really would happen
as a dominant disk grows.  The purpose of this article is to pursue this
question a little further.
 
It is well known that bars form in rotationally supported, massive disks that
lack a strong central density concentration.  Bars formed in this manner are
generally long lived, but they can be destroyed in at least two ways: by the
infall of a moderate mass companion into (or its passage through) the inner
disk (e.g., Pfenniger 1991; Athanassoula 1996) or, more interestingly,
through the growth of a central mass (Hasan \& Norman 1990; Norman, Sellwood
\& Hasan 1996).
 
Here we propose that every massive disk formed a bar in its early stages and
that many, though probably not all, of these bars were destroyed by the
development of a massive object in their centers.
 
It seems reasonable to expect that a newly formed bar in a young, gas-rich
disk will drive substantial inflow (Noguchi 1988; Barnes \& Hernquist 1991;
Friedli \& Benz 1993; Heller \& Shlosman 1994).  The distance from the center
at which the flow stalls depends both on where the quadrupole field of the
bar weakens and whether an inner Lindblad resonance (ILR) exists to halt the
flow (Athanassoula 1992).  Inflow is halted at an ILR because inside that
resonance the closed orbits in the potential are aligned perpendicular,
instead of parallel, to the bar (e.g.\ Binney \& Tremaine 1987, \S3.3).  The
torque from the bar on the gas, which arises from its offset distribution
with respect to the bar major axis, decreases rapidly within this resonance.
 
Since the halo has a large low density core, the disk which forms in the
early stages is likely to have a gently rising rotation curve and the bar
will lack an ILR.  Gas can therefore be driven as far inwards as the torques
from the bar can achieve.  It is difficult to predict what will happen when
large quantities of gas accumulate in a small volume at the center of a
galaxy, since it depends on the physical state of the gas, its ability to
fragment and form stars, energy feed-back, and other poorly-understood
processes.  For the purposes of this article we simply propose that a
gravitationally bound object is formed that is massive enough to weaken the
bar (\S3).  The precise nature of the object is unimportant for the dynamics
of the galaxy; a dense star cluster is one possibility, but it is also
natural to think that at least a fraction of the mass may collapse to create
the engine of a QSO, an idea we explore in section 5.
 
\section{Evidence for maximum disks}
The radial distribution of mass in a disk galaxy is strongly constrained by
its rotation curve.  The separate contributions from the individual stellar
populations and dark matter (DM) are not easily disentangled, however,
especially since there is generally no feature to indicate where the
component dominating the central attraction switches from luminous to dark
matter (Bahcall \& Casertano 1986).
 
\subsection{Rotation curve fitting}
Kalnajs (1983), Kent (1986), Buchhorn (1992), and Palunas \& Williams (1998)
all show that the shapes of the inner rotation curves of most (mostly high
surface brightness, HSB) galaxies are remarkably well predicted by the
visible matter if a constant M/L is assumed for the disk (and bulge).  In
these studies, mass discrepancies indicative of the DM halo contribution
become pronounced only in the outer parts.
 
Palunas \& Williams worked in the I-band, which Worthey (1994) finds is least
sensitive to metallicity.  Figure 1 shows the distribution of M/L$_{\rm I}$
values they obtained for their maximum disk fits, for $H_0=60$ km s$^{-1}$
Mpc$^{-1}$.  (Theirs was not a properly selected sample and statistical
inferences could therefore be misleading, but systematic effects in the
sample selection that would compromise this histogram seem unlikely.)  Some
galaxies were clearly barred, despite being classified as SA in the catalogs,
and the distribution of values for these cases is shown separately.  There is
no significant offset between the two distributions.
 
More than half their values lie in the range $1.5 < \hbox{M/L}_{\rm I} <
2.5$, although the distribution has broad tails on both sides.  (The spread
seems too great to be attributable to distance errors alone.)  Their values
therefore are consistent with the M/L$_{\rm I} \sim 1.5 -2$ predicted by
Jablonka \& Arimoto (1992) for continuous star formation and by Worthey
(1994) for a population with a mean age of $\sim 5$ Gyr.  Casertano \& van
Albada (1990) find that the M/L$_{\rm B}$ varies with color in the manner
predicted by earlier stellar population models.
 
If DM were to be important at all radii, its radial distribution would need
to be such that the shape of the inner rotation curve is little different
from that predicted by the light (van Albada \& Sancisi 1986; Freeman 1992),
presenting a fine-tuning problem for each galaxy.  Local features, such as
spiral arms, often reproduce small-scale structure in the rotation curve but
the stronger argument stems from the overall shape of the rotation curve. 
The strongest cases are the few known galaxies for which the rotation curve
declines somewhat outside the optical disk (e.g.\ Casertano \& van Gorkom
1991).  Nevertheless, the argument is not always regarded as compelling
(e.g.\ van der Kruit 1995).
 
\subsection{Barred galaxies}
Barred galaxies offer ways to estimate disk masses that are independent of
rotation curve fitting.  Weiner et al.\ (1998) model the gas kinematics in
the barred galaxy NGC 4123 by calculating 2-D gas flows in a model potential
derived from the I-band photometry.  They find that the strength of the
observed non-axisymmetric flow pattern can be matched only by making
the bar so massive that there is no slack in the rotation curve to permit
significant quantities of DM in the inner galaxy.  Their best fit model has
M/L$_{\rm I} = 2.0$ with values in the range $1.6\leq \hbox{M/L}_{\rm I} \leq
2.4$ being acceptable, in impressive concordance with Figure 1.
 
A second argument is presented by Debattista \& Sellwood (1998), who show
that bars are slowed dramatically as they lose angular momentum even in
moderate density halos.  Only if the halo central density is low, and the
disk is maximal, can rapid braking be avoided and pattern speeds remain
consistent with those in real barred galaxies.
 
\subsection{Global stability}
Ostriker \& Peebles (1973) suggested, in a frequently-cited argument against
maximum disks, that the global stability of unbarred disk galaxies requires a
massive halo.  Their original parameter, $t = T_{\rm rot}/|W|$, provides a
remarkably succinct summary of many results, both numerical and analytic, on
the stability of disks in potentials that have significant harmonic cores.
The criterion is clearly too simple for a number of reasons and several
attempts have been made to refine it (e.g., Christodoulou, Shlosman \&
Tohline 1995 and references therein).
 
Efstathiou, Lake \& Negroponte (1982, ELN) proposed another simple stability
criterion based on an extensive set of numerical simulations.  Once again,
their criterion requires substantial DM fractions to inhibit bars.
Unfortunately, it is now clear that the ELN criterion omits the important
stabilizing influence of a dense center.  Both linear stability work (Zang
1976; Toomre 1981; Evans \& Read 1998) and more careful $N$-body simulations
(Sellwood 1989 and this paper) show that linear bar-forming instabilities can
be avoided altogether in disks with minimal DM halos, provided only that the
rotation curve has a steep inner rise.  Toomre (1981) clearly states the
reason why a dense center is important and his argument is also explained by
Binney \& Tremaine (1987, ch.\ 6).  Sellwood (1989) was even able to identify
the numerical problem (excessive particle noise) that caused ELN to miss this
important factor.
 
In conclusion, no simple criterion has yet been formulated that encapsulates
all known global stability results but it is clear that any such criterion
cannot be based purely on the halo mass fraction.  Generally, unbarred
galaxies having gently rising rotation curves are still thought to require
substantial halo mass interior to the disk edge.  On the other hand, a
massive disk can be bar-stable even in a minimal halo when its rotation curve
rises steeply close to the disk center.
 
\subsection{Other Evidence}
Multi-arm spiral patterns develop in simulations in which the halo dominates
the central attraction everywhere, which Sellwood \& Carlberg (1984) argued
was consequence of swing amplification (Toomre 1981).  This local stability
property was exploited by Athanassoula, Bosma \& Papaioannou (1987) who
showed that the order of symmetry of the spiral arm pattern in many cases was
consistent with near maximum disks.
 
It is often argued that the Milky Way does not have a maximum disk (e.g.\
Kuijken 1995) and that if the maximum disk hypothesis fails for the Galaxy,
where we have the only real hope of a direct estimate of the disk density, it
is also unlikely to be valid where constraints are weaker.  Sackett (1997)
points out that this view is strongly sensitive to the adopted scale length
for the disk.  Furthermore, Sellwood (1998) and Englmaier \& Gerhard (1998)
suggest that the right M/L for models of the Milky Way based upon the COBE
NIR photometry again leaves little room for DM in the inner Galaxy.
 
Others claim to find evidence for large quantities of DM in the inner parts
of galaxies.  We find Bottema's (1993) arguments unconvincing because they
invoke general disk properties, such as a smooth exponential light profile
and/or a constant central surface brightness, etc.  The often substantial
departures from a simple exponential light distribution, which are usually
most pronounced in the massive inner disk, cannot be ignored.
 
\subsection{Implications}
Most of the above evidence is based upon bright, HSB galaxies and suggests
that in these cases DM halos have low central densities and large core radii.
 The rotation curves of low-luminosity galaxies (e.g.\ Broeils 1992) and of
low-surface brightness (LSB) galaxies (e.g.\ de Blok \& McGaugh 1996), on
the other hand, do not have the shape predicted from the light distribution
and therefore must have significant dark matter fractions right to their
centers.  The slowly rising rotation curves of these systems provide more
direct evidence of a large core to the halo mass distribution.
 
Navarro, Frenk \& White (1996, hereafter NFW) predict a cuspy density profile
for collisionless DM halos from their careful cosmological simulations, a
result first hinted at by Dubinski \& Carlberg (1991).  NFW concede that
their mass profile is inconsistent with the observed mass distributions in
low-luminosity galaxies (but see Kravtsov et al.\ 1998), and it also fails
for LSB galaxies (de Blok \& McGaugh 1996).  But Navarro (1997) claims both
that large HSB galaxies do have NFW halo profiles and that DM dominates the
mass even in the bright inner galaxy.  The evidence summarized above does not
support Navarro's claim, and it seems that few, if any, real galaxies have
the cuspy DM halo profile predicted by these large-scale structure
simulations.  It is unclear how this inconsistency will be resolved.
 
The dominance of luminous matter in the inner parts of large HSB galaxies
leads to the well-known conspiracy (Bahcall \& Casertano 1986) in which the
circular velocity from the luminous material in the inner parts is generally
differs little from that inferred for the dark matter in the outer parts.
Some galaxies with declining rotation curves are known (Casertano \& van
Gorkom 1991) but the drop is rarely more than 10\%.  Blumenthal et al.\
(1986) note that halo compression by baryon infall can lead to flat rotation
curves, but the similarity of the circular speeds even for extreme maximum
disks has no convincing explanation.
 
\section{Simulations}
In this section, we present simulations of disk galaxy formation in low
density halos.  The dynamics is largely controlled by the dominant mass
component, the stellar disk, but here we identify two gaseous processes which
are also of importance to the large-scale dynamics: dissipation to provide
fresh disk-like material on near circular orbits in the gravitational
potential well and strong shocks in bars to drive gas inwards and build up
large central mass concentrations.
 
Simulations of coupled gaseous and stellar dynamics present a much greater
computational challenge than do those of the stellar component alone.
Ambitious calculations to model both components at once have been presented
by Navarro \& Steinmetz (1997) and many others.  These expensive calculations
attempt to model not only the collisionless stellar component, but also gas
cooling, radiative heating, shocks, infall, star formation, energy feed-back
to the gaseous material, etc.  Many of these processes occur on scales too
small to be meaningfully resolved and the physics of star formation and
energy feed-back in particular is poorly understood; the simulators simply
include ad hoc rules to mimic them.  Moreover, the important process of
shock-driven inflow in bars is sensitive to numerical viscosity (Prendergast
1983) and therefore demands a high quality treatment.
 
Here we simplify the problem by including only those processes of importance
to the large-scale dynamics.  Sellwood \& Carlberg (1984), Carlberg \&
Freedman (1985), Villumsen \& Gunn (1987), and Toomre (1990) have established
the importance of dissipation in driving evolution in galaxy disks and have
shown that for dynamical purposes, all that is required is a steady reduction
of the rms peculiar velocity of particles to keep the system evolving.  Gas
inflow in bars has been shown to occur in both theoretical work (Prendergast
1983; Athanassoula 1992; etc.)\ and in practice (Quillen et al.\ 1995).
Other gas-dynamical processes have little effect on the mass distribution or
its kinematic properties, such as flow through spiral shocks and star
formation, and therefore scarcely affect the large-scale dynamics: settled
gas moves on essentially the same orbits as do stars since the generally mild
spiral shocks have a minor effect on the shape of the streamlines, while
conversion of gas into stars does not significantly change the orbit.  We
therefore employ a collisionless $N$-body code for our problem; the two
gaseous processes that are of dynamical importance have well understood
consequences which can be included in a simplified way (\S\S3.1 \& 3.2) while
other processes can be ignored.
 
Our models improve upon the work of Dalcanton et al.\ and Mo et al.\ in three
respects: (1) we follow the evolution of a galaxy as the disk mass rises,
allowing instabilities to develop and run their course, (2) we adopt a DM
distribution more consistent than the NFW profile with those observed, and
(3) we include the effect of a central massive object.
 
\subsection{Disk formation}
We mimic the gradual formation of a massive disk by adding fresh particles on
circular orbits in the mid-plane.  This crude procedure was originally found
effective by Sellwood \& Carlberg (1984).  Further experiments by Sellwood \&
Carlberg (unpublished) and Carlberg \& Freedman (1985) have demonstrated that
the formation of spiral disturbances, the build-up of random motion, and the
redistribution of angular momentum are scarcely affected by the details of
the procedure adopted to mimic the process of dissipation.
 
Large-scale structure simulations predict unrealistic halos and are saddled
with a celebrated ``overcooling problem'' (e.g.\ Navarro \& Benz 1991), which
is essentially that the specific angular momentum of the baryonic matter is
too low.  These difficulties suggest that such simulations cannot make
reliable predictions for the angular momentum distribution of infalling
material, or the rate at which it should arrive.  Moreover, energy feed-back
from star formation in the disk has an indeterminate effect on both
quantities.  Our rules for the accretion rate and for the initial angular
momenta of fresh particles are therefore necessarily arbitrary.
 
It seems reasonable to assume that the inner disk builds up quite rapidly and
that the mean angular momentum of late arriving material is greater than that
which was accreted at an early stage.  We have experimented with a few
different rules that embody both these principles and found that the final
mass distribution is not strongly sensitive to the accretion rate.  The mean
angular momentum of the accreted material does affect the outcome, but minor
changes to its distribution make little difference to the behavior.  The
rates at which we add mass are quite high in the early stages, as much 10\%
of the ``final'' mass of the disk per final rotation period in most cases; we
have experimented with decreasing this rate substantially during the later
stages.  The adopted rule for each of the two models reported here is given
in Table 1.
 
We give each added particle the local orbital velocity determined from
azimuthally averaged central attraction at the chosen radius and zero
velocity in the radial and vertical directions.  We determined the initial
$z$ coordinate from an estimate of the local mid-plane position, indicated by
where the vertical force passes through zero.
 
To save a little effort, we begin our calculations with a small
self-gravitating disk which we imagine to have formed at an early stage in
the center of a protogalaxy.  Our initial model has no strong density
concentration and no bulge.  In the experiments reported here, we adopt a
Kuz'min-Toomre model for our initial disk which has the radial surface
density profile $$
\Sigma(R) = {M \over 2\pi a^2} \left( 1 + {R^2 \over a^2} \right)^{-3/2}.
\eqno(1)
$$ We adopt the initial disk mass, $M$, and length scale, $a$, as our units
of mass and length.  We truncate the disk at $R=5a$, spread the particles
vertically with an rms thickness of $0.05a$, and assign suitable initial
velocities to the particles to create an equilibrium model with Toomre's $Q
\simeq 1.5$.  Wide variations around these choices make little difference to
the late time evolution of our models.  We evolve this initial disk until
after the central mass has reached its final value before beginning to add
fresh particles.
 
\subsection{Central mass}
As discussed above, a strong bar drives gas inward to produce a central mass
concentration.  We consider the possible fate of the dense gas concentration
produced by the inflow in \S5, but for the present purposes the only aspect
of relevance to the large-scale dynamics is the accumulation of mass in the
center.
 
Unreasonably high central masses ($\sim 5\%$ of the disk mass) are needed to
destroy the bar completely and are unlikely to be achieved in reality.  Gas
inflow is halted at the ILR, which is certainly present once the central mass
has reached a mass of $\sim 1-2\%$, cutting off the supply of material for
further growth.  We therefore limit the mass of the central object to $\sim
1.5$\% of the initial disk mass.  This limit is sufficient to cause a
significant weakening of the bar, to about half the amplitude which results
if no central mass is imposed.
 
We increase the mass of a single central particle having the density profile
of a Plummer sphere.  The effect the central mass has on the bar depends to
some extent on its core radius: For a fixed final mass, we find that the bar
is weakened to a much lesser extent when its core radius exceeds $0.05a$ but
reducing it below this value causes the bar to weaken only slightly more, as
might be expected.  We were unwilling to employ a very small core radius
since that would make the calculations more expensive by requiring a shorter
time step.
 
In the two experiments reported here, we set the core radius to $0.05a$ and
increase the mass of the central particle whenever the ratio $\alpha_2 /
\alpha_0$ exceeds some threshold value.  Here the $\alpha$s are coefficients
of a Fourier expansion of the density over some inner radial range.  We
choose these parameters so as to obtain a final mass of $\sim 1.5\%$; the
adopted values for each model are given in Table 1.
 
\subsection{Halo}
Since we were anxious to model maximum disks, and wished to avoid including
rigid potentials if possible, our first models lacked a halo component.  This
seemed justifiable, since most halo mass in a maximum disk model lies beyond
the disk edge and, when spherically distributed, exerts no forces on the
disk.  We found, however, that irrespective of the amount of mass added to
the outer disk, the rotation curve always declined in our halo-less models.
The massive disk always developed strong one- and two-armed spiral patterns
that caused the central density to rise ever higher while spreading material
to ever larger radii, thereby defeating our objective of creating a realistic
rotation curve.\footnote{Note that such behavior challenges suggestions that
the flat rotation curves of galaxies could be caused by an outwardly
increasing M/L in the disk or by the existence of large quantities of cold
gas in the outer disks (Pfenniger, Combes \& Martinet 1994).}
 
We were therefore forced to include a DM halo in some way.  We add
supplemental forces from a rigid potential of the form $$
\Phi_{\rm halo} = {V_0^2 \over 2} \ln \left( 1 + {r^2 \over c^2} \right),
\eqno(2)
$$ which yields an asymptotically flat circular velocity of $V_0$ for $r \gg
c$, with $c$ being the ``core radius''.  Setting $c=30a$, 2.5 times the
largest radius at which we added particles, reduced the spreading of the
outer disk dramatically.  Since the peak circular velocity from our disk is
about $0.6 (GM/a)^{1/2}$, we generally choose $0.6 \leq V_0(a/GM)^{1/2} \leq
0.8$; our results are little affected by the precise value within this narrow
range.
 
We include this diffuse halo as a rigid component because to represent it
with live particles would severely compromise our spatial resolution in the
disk.  By keeping it rigid, we introduce a number of approximations, which we
do not expect to affect our conclusions.  First, we exclude the possibility
of disk-halo interactions, such as angular momentum exchange (Weinberg 1985)
but this is unlikely to be significant for such a large core radius halo
(Debattista \& Sellwood 1998).  Second, we exclude the compression of the
halo as the disk mass builds (Barnes \& White 1984).  Third, a substantial
lop-sidedness in the distribution of active particles could also give rise to
unphysical behavior, but has fortunately not developed in any of our models
with halos.  This final concern forced us to fix the position of the central
mass also.
 
\subsection{Numerical details}
We use the 3-D polar grid-based $N$-body code described by Sellwood \&
Valluri (1997).  The numerical parameters for the models presented here are
given in Table 1.  Reasonable variations around the adopted values do not
lead to significantly different behavior.
 
\subsection{Results}
We first describe the results from an experiment (model 1) which is typical
of our more successful models.  All quantities from here on are expressed in
units such that $G=M=a=1$.
 
As expected, the initial massive disk rapidly forms a bar in the usual way.
Because the model is not centrally condensed, the pattern speed need not be
very high to avoid an ILR, as shown in Figure 2(a).
 
The central density of the model rises in response to the growing central
mass making a small bulge of much larger mass than that of the imposed mass.
The bar amplitude, assessed as $\alpha_2 / \alpha_0$, abruptly drops to about
half its peak as the central mass reaches 1.4\% of the disk mass in this
case.  At this point, the mass distribution is sufficiently centrally
condensed that the bar has probably acquired an ILR, as suggested in Figure
2(b).  It should be noted, however, that the curves of $\Omega - \kappa/2$
drawn in this Figure are computed from the azimuthal average of the central
force in this strongly barred potential.  They do not, therefore, give a {\it
reliable\/} guide to the existence of the perpendicularly aligned orbit
family, which is the only sure indication of a generalized Lindblad resonance
in strongly perturbed potentials (Contopoulos \& Grosb\o l 1989).
 
Once we stop growing the central mass, the rotation curve of the model
(Figure 3) has a high inner peak and a deep dip before rising again as the
halo contribution picks up.  We began to add fresh particles at a steady
rate, from time 160, placing them on circular orbits in the radial range
$8<R/a<12$ only, in line with our expectation that later arriving material
will have somewhat larger angular momentum.  We added 8 particles at every
time step, or $\dot M = 5\times 10^{-3}$ in our units -- i.e.\ a mass equal
to one tenth the final disk mass about every rotation period at $R=10$.
Strong spiral patterns develop (Figure 4) which redistribute angular momentum
efficiently.  Continued accretion of material causes recurring spiral
activity that spreads the new material both inwards and outwards, causing the
rotation curve to become more nearly flat while the inner peak rises slightly
more.
 
The residual bar left after the growth of the central mass is short and weak.
 Later in this simulation it disappears almost entirely (Figure 4b), and no
bar-like feature returns.  The gradual disappearance of the bar seems to be
caused by an interactions with strong spiral patterns in the inner disk.
 
By the last time shown in Figure 4, which is not the end of our calculation,
the accreted mass is four times that of the initial disk.  Even though the
disk created in this way is almost fully self-gravitating, it does not form a
bar.  The instability is inhibited both because the inner disk is dynamically
hot and because of inner Lindblad resonances arising from the high central
density (Zang 1976; Toomre 1981; Binney \& Tremaine 1987 \S6.3; Sellwood
1989; Evans \& Read 1998).  It is worth noting that the $t$ stability
parameter introduced by Ostriker \& Peebles (1973) remains above 0.25 from
time 300 to the end; the stability parameter introduced by Efstathiou et al.\
(1982) remains $\ltsim0.7$ since the disk scale length increases
approximately as the disk mass.
 
A notable feature of this experiment is that the matter rearranges itself in
such a way that the rotation curve becomes approximately flat, except perhaps
for the sharp inner peak (Figure 3).  We have seen this behavior in other
experiments with different accretion rules.  Moreover, despite concerted
attempts, we have been unable to create a persisting rise in the rotation
curve in a massive disk; whenever accretion created a rising rotation curve,
very strong spiral patterns developed that redistributed enough angular
momentum to cause the central peak to rise until a roughly flat or declining
rotation curve was re-established.  This result seems to be of some
importance.
 
This experiment suggests a natural mechanism to build massive unbarred disks
that have central densities high enough to suppress bar formation and a
roughly flat rotation curve.  It demolishes the argument by Mo et al.\ that
self-gravitating disks could not be formed, and supports the contention by
Dalcanton et al.\ that instabilities will form a modest bulge at the centers
of disk dominated galaxies.
 
The galaxy we have created has a rotation curve (e.g.\ Rubin, Ford \&
Thonnard 1980), bulge size and morphological appearance resembling those of
large Sc galaxies.  (Indeed, Kent (1986) showed that the inner parts of
precisely these same rotation curves were nicely reproduced by the radial
light distribution in these galaxies.)  We have produced an Sc type galaxy
probably because our simulation has mimicked the formation of an isolated
galaxy.  Early-type galaxies may require mergers of proto-galactic fragments,
or even of small disk galaxies, to make their larger bulges.
 
\section{Barred galaxies}
Since we argue that most massive disks acquire central masses which weaken or
destroy their bars, we must account for the existence of a substantial
fraction of strongly barred galaxies today.  Before offering an explanation,
we note two other related, but long-unanswered, questions presented by barred
galaxies.
 
\begin{itemize}
 
\item The observational signature of a central density high enough that it
should have inhibited a bar by Toomre's (1981) mechanism is the existence of
a nuclear stellar or gaseous ring.  Many barred galaxies have nuclear rings
with diameters of several hundred parsecs (Buta \& Crocker 1993), which are
widely believed to form where gas inflow along the bar is halted at an inner
Lindblad resonance (ILR).  There is abundant evidence for gas build up in
nuclear rings (e.g.\ Helfer \& Blitz 1995; Rubin, Kenney \& Young 1997) and
other evidence for ILR phenomena has been claimed (e.g.\ Knapen et al.\
1995).  We note also that the Milky Way, which is now believed to be a barred
galaxy (de Vaucouleurs 1964; Blitz et al.\ 1993), has long been known to have
a very dense center (Becklin \& Neugebauer 1968).  So why are these galaxies
barred, if their dense centers should have inhibited bar formation?
 
\item No satisfactory explanation for the observed frequency of bars has yet
been proposed.  Some 30\% of HSB galaxies are strongly barred and perhaps an
equal fraction are weakly barred (Sellwood \& Wilkinson 1993).  An argument
that these galaxies contain somewhat less DM than do unbarred galaxies is
easily dismissed by the evidence for maximum disks in all HSB galaxies
presented in section 2.  Moreover, while the normal bar instability provides
a natural explanation for the existence of strong bars, we are not aware of a
model for the origin of weak bars.
 
\end{itemize}
 
In our picture, most HSB galaxies should have dense centers -- barred
galaxies should not be an exception.  Here we propose that a second, and this
time lasting, large-scale bar developed in some galaxies.  Unlike the first
bar, it is not weakened by gas inflow because the flow is halted at the ILR
which prevents the central mass from increasing.
 
A second bar could form for one of two possible reasons.  First, the
stability of the disk relies on the ability of the ILR to damp incipient bar
instabilities, but the resonance can be overwhelmed by large perturbations
and a strong bar can result (Sellwood 1989).  One possible source of large
perturbations is tidal encounters (Noguchi 1987); evidence has been claimed
(Elmegreen, Elmegreen \& Bellin 1990) for a higher fraction of barred
galaxies in dense environments where bar-triggering encounters should be more
frequent.
 
Second, we should expect relatively low angular momentum material to be added
to the disk in some fraction of galaxies.  Such material will settle in the
inner few kpc where it is able to re-excite a bar, as shown in Figure 5.
(This model differs in several respects from that shown in Figure 4.  The
central mass was grown to about 1.5\% of the disk mass much more quickly and
accretion began at time 40.  Note also that the spatial scale in the plots is
different.)  The short initial bar, which was weakened when the central mass
formed, gradually regains strength through angular momentum exchange with
fresh cool material near its outer Lindblad resonance, as occurred in
previous experiments (Sellwood 1981).  The distribution of angular momentum
in the accreted matter will need to be known in some detail before it can be
determined whether the observed distribution of bar strengths is correctly
predicted by this effect.
 
In this picture, we expect the fraction of galaxies possessing strong bars to
be similar, or perhaps somewhat less, at high redshift than in the local
universe.  The timescale on which the first bars are formed and destroyed is
so short that we are unlikely to see many galaxies in this stage.  The
formation of the second bar occurs over a longer period, though still in the
early life of a galaxy, and we therefore might expect to see slightly fewer
in a high redshift sample.  It is important that the frequency of bars in the
high redshift sample should be determined from images in the rest-frame red
or near-IR pass bands in which bars are most easily seen in the nearby
universe.
 
\section{The quasar epoch}
The now standard model for QSO activity is a massive collapsed object at the
center of a galaxy, proposed by Lynden-Bell (1969).  Mass constraints require
that such objects are active for a comparatively short time (Padovani, Burg
\& Edelson 1990).  Small \& Blandford (1992) argue, from the existence of
high redshift quasars, that the massive object must grow ``rapaciously'' in
its early stages.  Quasar spectra seem to require massive objects in most
bright galaxies (Chokshi \& Turner 1992).
 
We suggest that some of the mass which accumulates at the centers of bars
collapses to power quasars.  If this is association is correct, the onset of
the quasar epoch (Schmidt, Schneider \& Gunn 1995; Shaver et al.\ 1996)
should be coincident with the formation of the first bars in disks.
 
The idea that large-scale stellar bars could cause massive objects to be
created was already proposed by Shlosman, Begelman \& Frank (1990), who
suggested that the gas density could rise to the point of a second, much
smaller scale, bar instability -- or indeed a cascade of such events to
successively smaller spatial and time scales.  Haehnelt \& Rees (1993) and
Eisenstein \& Loeb (1995) also suggested that massive objects formed in the
centers of gas rich protogalaxies, but these models do not invoke
non-axisymmetric stellar mass distributions to remove the angular momentum.
 
The new feature of our model is that it offers a dynamical reason why nuclear
activity should cease soon after the central mass becomes large enough to cut
off its gas supply through the development of an inner Lindblad resonance, or
(possibly) the complete destruction of the bar.  The later formation of bars
in a significant fraction of galaxies would not cause further nuclear
activity because the central engine is shielded by the ILR.
 
QSOs should flare during the formation of disk galaxies, but we also expect
that mergers of disk galaxies which already host central masses will lead to
brighter outbursts (Lehnert et al.\ 1992) as the mass of the central object
rises further.  This seems to be required by the observation that QSOs,
especially the radio-loud type, are frequently found in elliptical galaxies
(Taylor et al.\ 1996).  Mergers lead to tri-axial objects (Barnes 1992), and
Merritt \& Quinlan (1998) have already suggested that if the mass of the
central object in a tri-axial elliptical galaxy is low, it will continue to
be fed by stars on box orbits until it reaches a mass approaching 2\% of the
galaxy before the box orbits become stochastic and make the galaxy
axisymmetric.
 
\subsection{Massive central objects today}
Our model suggests that supermassive objects should be found in almost every
bright galaxy, except for those cases where they have possibly been ejected
(Rees 1997).  Magorrian et al.\ (1997) claim that the data on nearby galaxies
are consistent with a massive object in almost every galaxy, with a
suggestion that the mass of the central object is correlated with the bulge
mass.  Both aspects can be explained in our picture, since the dissolved bar
produces a bulge.  Their estimate of an object mass that is a few tenths of a
percent of the bulge mass may suggest that perhaps the supermassive object
may be some $\sim10$\% of the central mass concentration.  It should be noted
that disks will continue to increase in mass through accretion subsequent to
this event.  Minor mergers will also add mass, probably to the bulge.
 
We predict that no massive central objects should be found in DM dominated
galaxies, because such galaxies would not have formed bars.  In section 2 we
argue that HSB galaxies are disk dominated, but it is well established that
LSB galaxies cannot be (de Blok \& McGaugh 1996).  There is also some
evidence for significant halos in low luminosity galaxies (e.g.\ Broeils
1992); a good local example is M33 which has a gently rising rotation curve
and must therefore be halo dominated to suppress the bar instability.
Happily, M33 also happens to have one of the lowest upper limits on the mass
of a central object (Kormendy \& McClure 1993).  It should be noted that this
difference is not predicted by models which invoke capture of pre-formed
black holes into the centers of galaxies (e.g., Lacey \& Ostriker 1985) or in
purely gas dynamical models (Eisenstein \& Loeb 1995; Silk \& Rees 1998) in
which the processes invoked do not depend on DM fraction.
 
\section{Loose ends}
While we hope the outline for the late stages of disk galaxy formation
presented in this paper has some merit, we recognise that many crucial parts
of the picture are seriously incomplete.  More detailed examination of
several parts may render the whole edifice untenable.  Here we note several
significant gaps of which we are aware.
 
We have simply postulated that a dense central object forms in a gas rich
barred galaxy which originally had a shallow density distribution.  Most
bright galaxies have dense centers today, and it seems reasonable that both
dissipation and angular momentum redistribution were required to bring this
about.  Gas inflow along a bar is well established, but it is highly
speculative to argue, as we have here, that a dense central object is built
where there was not one before.
 
The inflow process is self-regulating, since it is halted by the ILR as soon
as enough mass has accumulated in the center to produce one.  The existence
of an ILR in a strongly barred system cannot simply be determined from the
``rotation curve'' but requires the perpendicularly aligned orbits in the bar
to be found (e.g.\ Contopoulos \& Grosb\o l 1989).  Exactly when this family
appears and how much mass can be expected to make it to the center before the
valve closes is not yet known, but must be of the order of 1\%.  Further work
in this area is needed.
 
\Ignore{
What is the distribution of angular momentum in the infalling gas?  Spergel
(private communication) points out that if the overdense region in which the
galaxy formed were perfectly spherical with each shell being in uniform
rotation, then a significant fraction of material should have very little
angular momentum, which might lead too large a fraction of barred galaxies.
No forming galaxy is completely isolated, of course, and the infalling
gaseous material may well have begun to condense into sub-clumps, but it is
far from clear that the fraction of very low angular momentum material could
be as small as we require.}
 
Can the currently observed fractions of strong and weakly barred galaxies be
accounted for naturally by a combination of interactions and accretion of
low-angular momentum material?  A quantitative answer to this question could
be provided only from simulations of hierarchical clustering, but seems well
beyond what is technically feasible today.
 
We argue that our models resemble large Sc galaxies both in appearance and in
their rotation curve shapes.  Lower luminosity Sc galaxies with gently rising
rotation curves must have sufficient DM to inhibit the bar instability.  It
is likely that the formation of early type galaxies with dominant bulges
requires some degree of merging of protogalactic fragments which has to
happen at a time after the dense center is established in at least one of the
fragments but while more infalling material can create a new disk.  This
process will also have to be modelled carefully to determine how our model is
affected.
 
We have used an unresponsive mass distribution to represent the halo, an
approximation with the limitations we listed in \S3.3.  It is clearly
desirable to simulate a live halo to show that some initial halo mass
distribution can be adopted that will allow a maximum disk to form within it.
 
\Ignore{Here we have modelled accretion in a highly simplified manner by
adding fresh particles to the disk on circular orbits.  We deliberately
adopted this approach in order to avoid the complicated gas dynamics of
dissipative infall and star formation.  It seems unlikely to us that a that a
more realistic treatment would cause a significantly different outcome, but
it should be checked.}
 
\section{Conclusions}
We have extended the popular picture of disk galaxy formation through the
settling of gas in DM halos to address a variety of unsolved problems.  We
suggest that: (1) Not only can massive disks survive at the centers of low
density halos, but that they naturally develop rotation curves which become
flat over a wide radial range after a steep inner rise.  (2) The process of
developing the dense central concentration, and much more speculatively the
central engine of quasars, occurs when the bar, which must form in every
massive disk, drives gas inwards until the central object is massive enough
to cut off the flow.  (3) This amount of mass is sufficient to weaken the
bar, though not to destroy it.  (4) Galaxies with central mass concentrations
can be barred today because either they suffered a tidal encounter or they
later accreted low-angular momentum material that strengthened the bar.  The
second bar in a galaxy can survive, without causing the central mass to rise
further or re-igniting the central engine, because inflow this time is halted
at the inner ring which forms at the ILR.
 
If these ideas are correct, then we predict that massive objects should not
be found in galaxies in which DM dominates all the way to the center.
 
This entire picture is speculative and requires a great deal of additional
work to establish its viability.  We list some of the more important loose
ends in section 6.
 
\bigskip
This work was completed while JAS was a visitor to the Isaac Newton Institute
in Cambridge; their support is greatly appreciated.  This work was supported
by NSF grants AST 93/18617 and AST 96/17088 and by NASA Theory grant NAG
5-2803.  Conversations with David Merritt and Stacy McGaugh have been very
useful.  Ray Carlberg, Jeremy Goodman, Martin Haehnelt, Jim Pringle, Scott
Tremaine and the referee, Alar Toomre, made helpful comments on the manuscript.

\vfill\eject

\newcount\style
\style=1

\begin{figure}[t]
\ifodd\style
  \centerline{\psfig{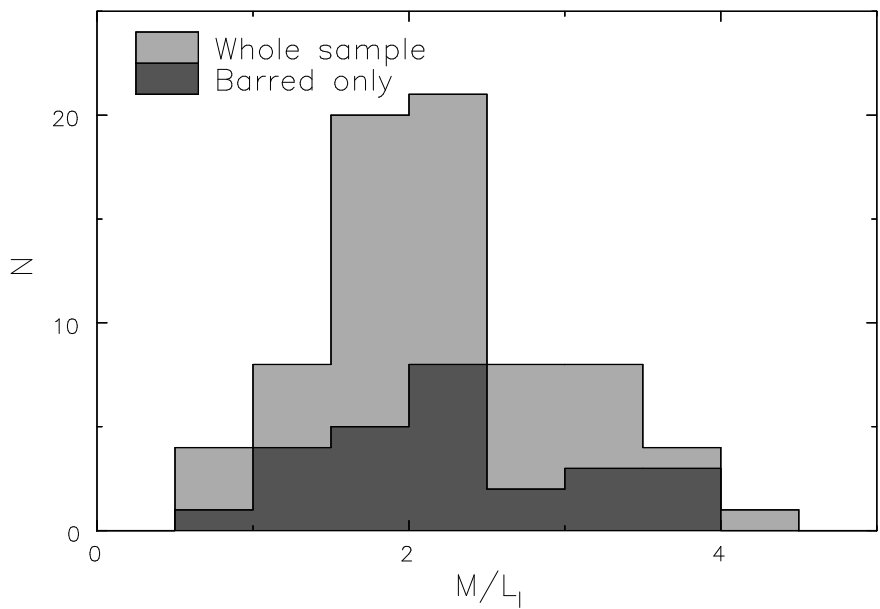}}
\fi
 
\caption{M/L$_{\rm I}$ ratios obtained by Palunas \& Williams (1998) for the
disk components in their maximum disk fits.  Their values have been adjusted
for $H_0=60$ km s$^{-1}$ Mpc$^{-1}$.}
 
\end{figure}
 
\begin{figure}[t]
\ifodd\style
  \centerline{\psfig{figure=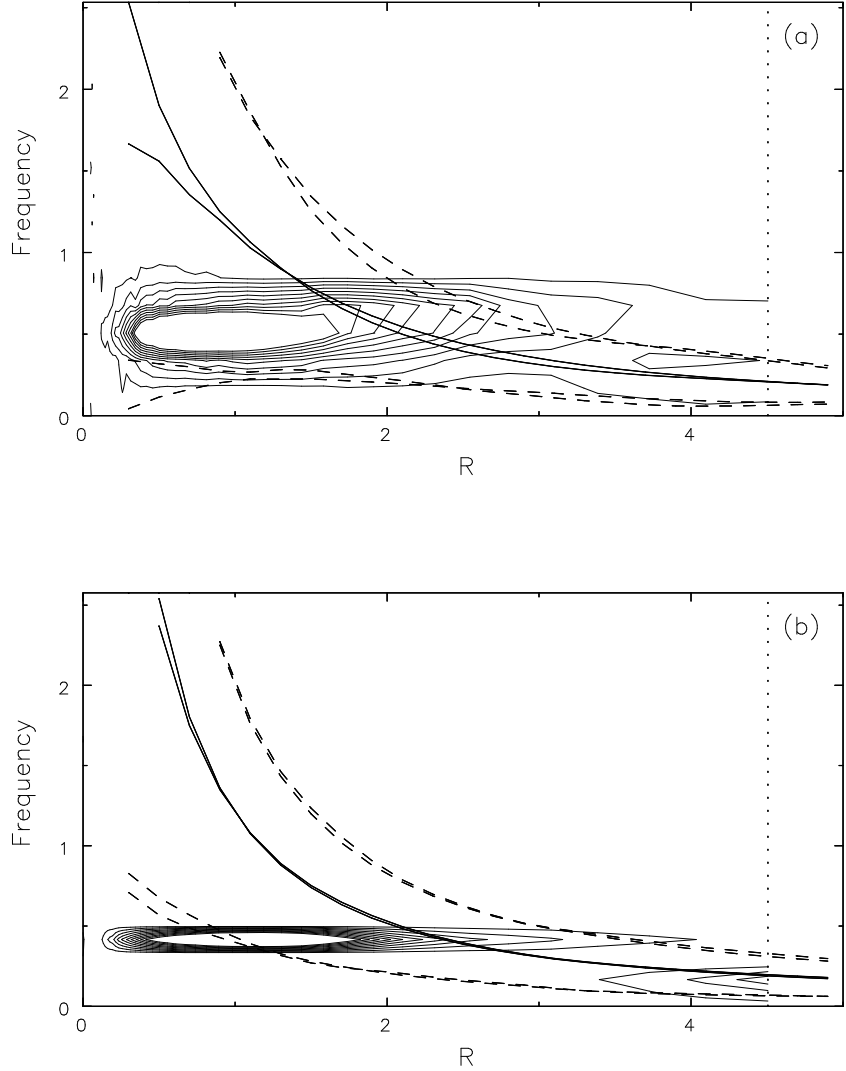,width=0.7\hsize,clip=,angle=0}}
\fi

\caption{Contours of power in bi-symmetric density perturbations as a
function of radius and frequency over two different time intervals in model
1. The upper plot is over a short initial period ($0\leq t \leq 40$) during
which the bar forms, the lower plot is from a later time interval $(95\leq
t\leq170$) as the central mass reaches its final mass.  The full-drawn lines
show curves of the circular angular frequency $\Omega$ and the dashed curves
mark $\Omega\pm\kappa/2$.  In both plots, these curves are determined from
the mean central attraction in the model at the beginning and end of the
adopted time range.}
 
\end{figure}
 
\begin{figure}[t]
\ifodd\style
  \centerline{\psfig{figure=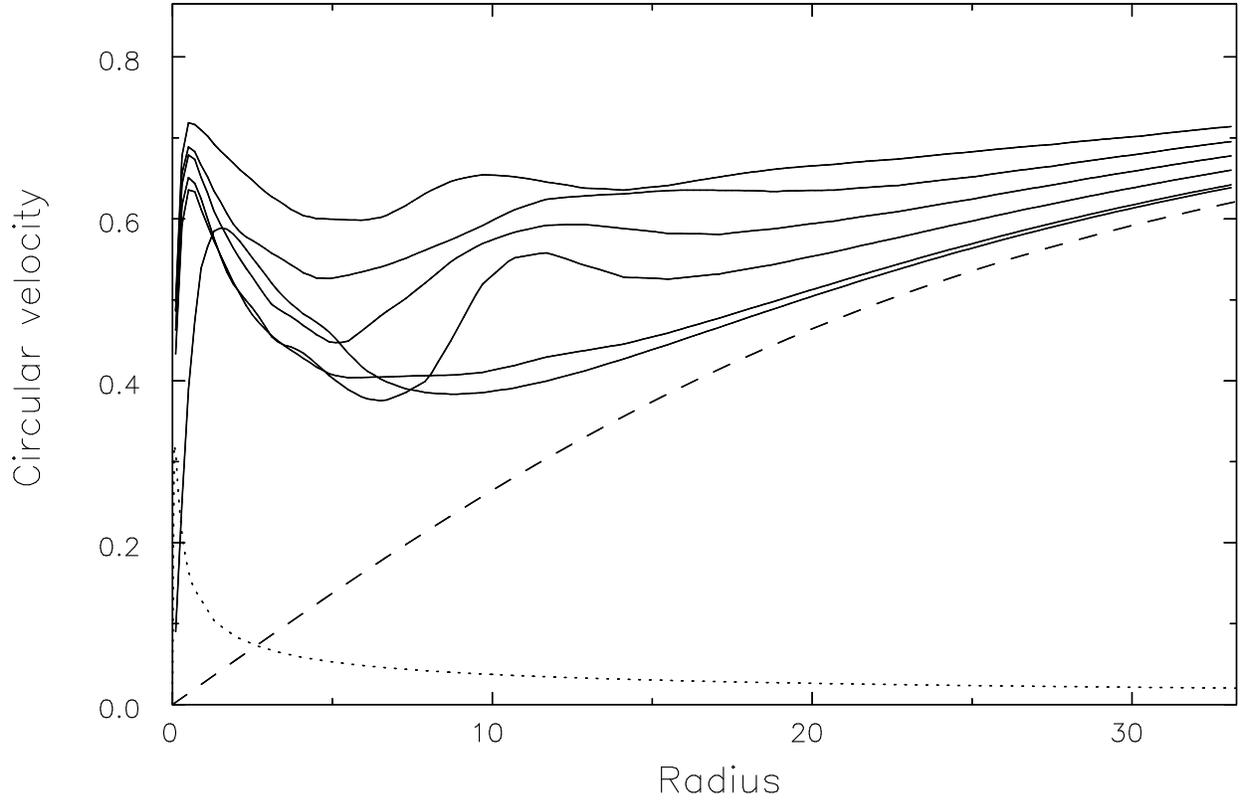,width=\hsize,clip=,angle=0}}
\fi
 
\caption{The rotation curve at the six times 0(192)960 (full-drawn lines);
the circular velocity at both $R\sim1$ and $R\sim10$ rises monotonically over
time.  The dashed curve is the fixed halo contribution and the dotted curve
shows the contribution from the central mass only, which is absent at $t=0$
but the same for all other times shown.}
 
\end{figure}
 
\begin{figure}[t]
\ifodd\style
  \centerline{\psfig{figure=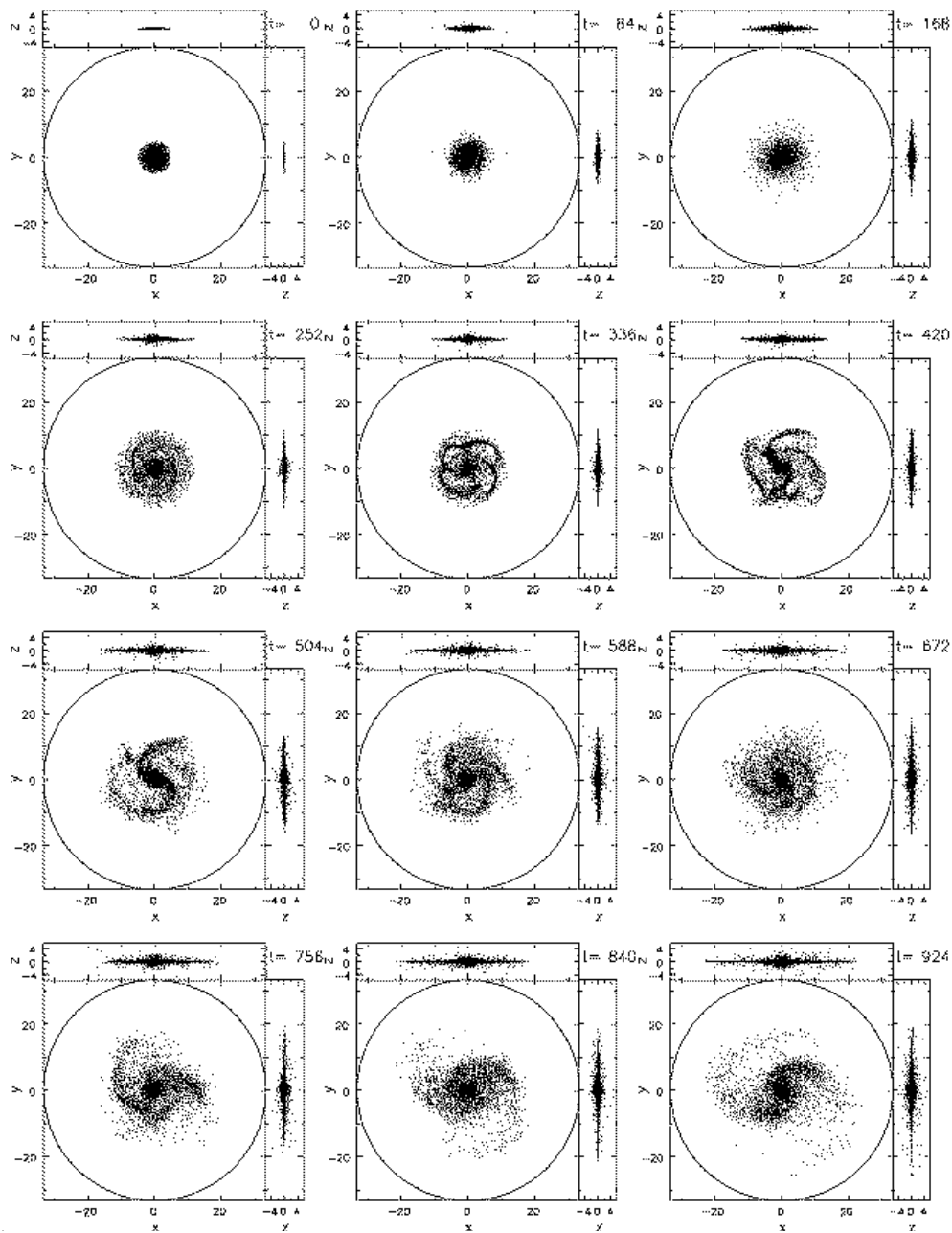,width=0.8\hsize,clip=,angle=0}}
\fi
 
\caption{(a) Snapshots of model 1 at equally spaced times.  The number of
particles in the simulation increases after time 160, but each plot shows a
random selection of about 5000.  Particles are added in the range $8 \leq R
\leq 12$ only, but are spread radially by the strong spiral patterns.  Notice
that this almost fully self-gravitating disk has no bar in the second half of
the evolution.}
 
\end{figure}
 
\setcounter{figure}{3}
\begin{figure}[t]
\ifodd\style
  \centerline{\psfig{figure=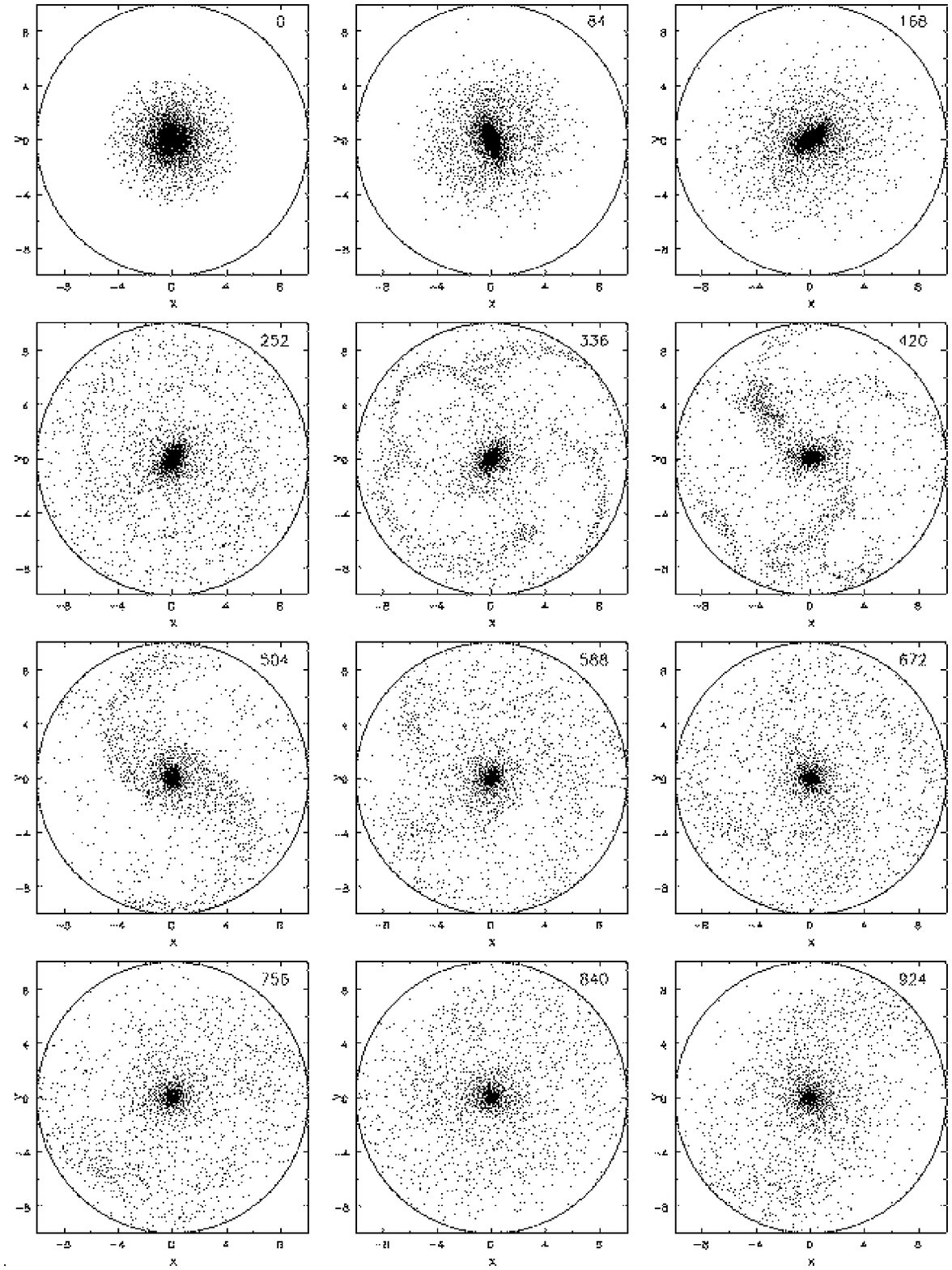,width=0.8\hsize,clip=,angle=0}}
\fi

\caption{(b) Close-up views of the inner region of the model shown in (a).}
 
\end{figure}
 
\begin{figure}[t]
\ifodd\style
  \centerline{\psfig{figure=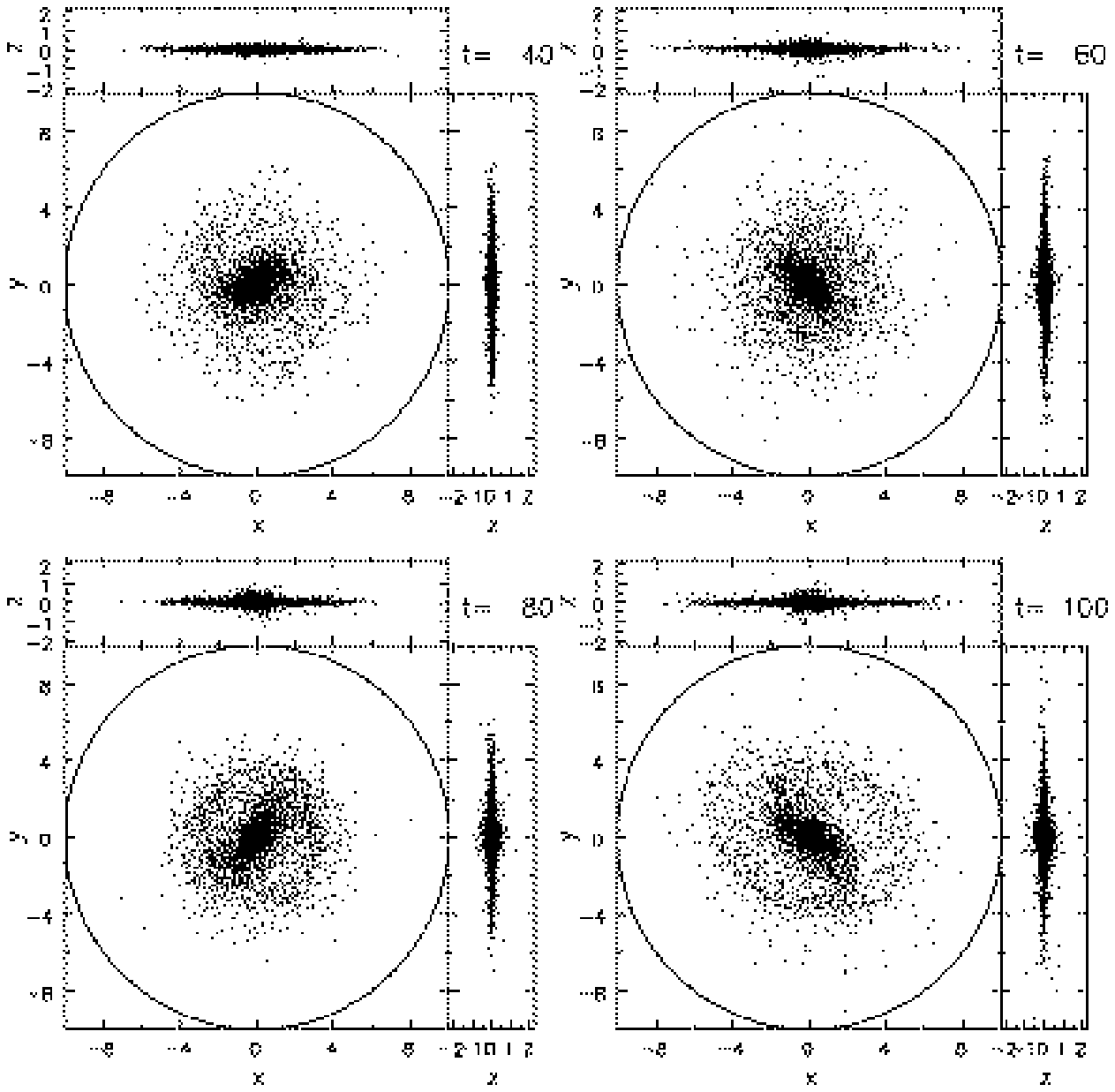,width=\hsize,clip=,angle=0}}
\fi
 
\caption{Snapshots of model 2 in which particles are added at much smaller
radii than in model 1.  This low angular momentum material causes the bar to
gain strength, unlike in Figure 4.}
 
\end{figure}
 
\vfill\eject
 
\begin{table}[t]
\setlength{\baselineskip}{12pt}
\centering
\begin{tabular}{lccc}
   \multicolumn{4}{c}{Table 1.
    Summary of the two models} \\ \hline \hline
   \multicolumn{1}{c}{Quantity} &
   \multicolumn{1}{c}{Model 1}  &
   \multicolumn{1}{c}{Model 2} & \multicolumn{1}{c}{unit} \\ \hline
   \multicolumn{4}{c}{Numerical Parameters} \\ \hline
   Initial number of particles \dotfill & $4 \times 10^4$ & $4 \times 10^4$ \\
   Grid size $(R,\phi,z)$ \dotfill & $72\times66\times375$  &
$65\times64\times225$  \\
   Vertical plane spacing \dotfill & $0.03$ & $0.02$ & $a$ \\
   Grid boundaries $(R,z)$ \dotfill & $(33.6,\pm5.61)$ & $(20.1,\pm2.24)$ &
$a$ \\
   Softening length \dotfill & $0.05$ & $0.05$ & $a$ \\
   Time step \dotfill & $0.04$ & $0.02$ & $(a^3/GM)^{1/2}$ \\ \hline
   \multicolumn{4}{c}{Initial Disk} \\ \hline
%   Model \dotfill & Kuz'min-Toomre  & Kuz'min-Toomre \\
%   Mass \dotfill & $M$ &  $M$  \\
   Toomre $Q$ \dotfill & $1.5$ & $1.5$  \\
   RMS vertical thickness \dotfill & $0.05$ & $0.05$ & $a$ \\
   Truncation radius \dotfill & $5$  & $5$ & $a$ \\ \hline
   \multicolumn{4}{c}{Accretion rule} \\ \hline
   Time accretion starts \dotfill & $160$  & $40$ & $(a^3/GM)^{1/2}$ \\
   Particles added per time step \dotfill & 8  & 4  \\
   Accretion rule \dotfill & uniform in $R$ & Gaussian in $J$ \\
   & $8\leq R \leq 12$ & $\bar J= 1+0.01t$ $\sigma=0.5$ & $a$, $(GMa)^{1/2}$
\\  \hline
   \multicolumn{4}{c}{Fixed halo} \\ \hline
%   Model \dotfill  & Isothermal with core & Isothermal with core \\
   $V_0$ \dotfill &  $0.7$ &  $0.7$ & $(GM/a)^{1/2}$ \\
   Core radius \dotfill & $30$ & $30$ & $a$ \\ \hline
   \multicolumn{4}{c}{Central Mass} \\ \hline
%   Model \dotfill  & rigid Plummer sphere & rigid Plummer sphere \\
   Core radius \dotfill & $0.05$  & $0.05$ & $a$ \\
   Radial range for $\alpha_2/\alpha_0$ \dotfill & $1\leq R \leq 2$ & $1\leq
R \leq 2$ & $a$ \\
   Mass rises when $\alpha_2/\alpha_0 >$ \dotfill & 0.4 & 0.4 \\
   $dM/dt$ \dotfill & $2.5\times10^{-4}$ & $2.5\times10^{-3}$ &
$(GM^3/a^3)^{1/2}$ \\
   Period of growth \dotfill & $24\leq t \leq 80$ & $24\leq t \leq 30$ &
$(a^3/GM)^{1/2}$ \\
   Final mass \dotfill & $0.014$ & $0.017$ & $M$ \\  \hline \hline
\label{tab:one}
\end{tabular}
\end{table}
 

\begin{references}
\reference{} Athanassoula, E.\ 1992, \mnras, {\bf 259}, 345
 
Athanassoula, E.\ 1996, in {\it Barred Galaxies}, IAU Colloq.\ 157, ed.\ R.\
Buta, D.\ A.\ Crocker, \& B.\ G.\ Elmegreen (ASP Conf series 91), 309
 
Athanassoula, E., Bosma, A.\ \& Papaioannou, S.\ 1987, \aap, {\bf 179}, 23
 
Bahcall, J.\ N.\ \& Casertano, S.\ 1986, \apj, {\bf 308}, 347.
 
Barnes, J.\ E.\ 1992, \apj, {\bf 393}, 484
 
Barnes, J.\ E.\ \& Hernquist, L.\ E., 1991, \apj, {\bf 370}, L65
 
Barnes, J.\ \& White, S.\ D.\ M.\ 1984, \mnras, {\bf 211}, 753.
 
Becklin, E.\ E.\ \& Neugebauer, G.\ 1968, \apj, {\bf 151}, 145.
 
Binney, J.\ \& Tremaine, S.\ 1987, {\it Galactic Dynamics}, (Princeton:
Princeton University Press)
 
Blitz, L., Binney, J., Lo, K.\ Y., Bally, J. \& Ho, P.\ T.\ P.\ 1993, Nature,
{\bf 361}, 417
 
Blumenthal, G.\ R., Faber, S.\ M., Flores, R.\ \& Primack, J.\ R.\ 1986,
\apj, {\bf 301}, 27
 
Bottema, R.\ 1993, \aap, {\bf 275}, 16
 
Broeils, A.\ 1992, PhD Thesis, Groningen University
 
Buchhorn, M.\ 1992. PhD Thesis, Australian National University
 
Buta, R.\ \& Crocker, D.\ A.\ 1993, \aj, {\bf 105}, 1344
 
Carlberg, R.\ G.\ \& Freedman, W.\ L.\ 1985, \apj, {\bf 298}, 486
 
Casertano, S.\ \& van Albada, T.\ S., 1990, In {\it Baryonic Dark Matter},
eds D.\ Lynden-Bell \& G.\ Gilmore (Dordrecht: Kluwer) p~159
 
Casertano, S.\ \& van Gorkom, J.\ H., 1991, \aj, {\bf 101}, 1231
 
Chokshi, A.\ \& Turner, E.\ L., 1992, \mnras, {\bf 259}, 421
 
Christodoulou, D.\ M., Shlosman, I.\ \& Tohline, J.\ E.\ 1995, \apj, {\bf
443}, 551
 
Contopoulos, G.\ \& Grosb\o l, P., 1989, \aap\ Rev., {\bf 1}, 261
 
Courteau, S.\ \& Rix, H-W.\ 1997, astro-ph/9707290
 
Dalcanton, J., Spergel, D.\ N.\ \& Summers, J.\ J.\ 1997, \apj, {\bf 482}, 659
 
Debattista, V.\ P.\ \& Sellwood, J.\ A.\ 1998, \apj, {\bf 493}, L5
 
de Blok, W.\ J.\ G., \& McGaugh, S.\ S.\ 1996, \apjl, {\bf 469}, L89
 
de Vaucouleurs, G.\ 1964, in {\it The Galaxy and the Magellanic Clouds\/} IAU
Symposium {\bf 20}, eds.\ F.\ J.\ Kerr \& A.\ W.\ Rodgers (Australian Academy
of Science, Sidney) p~195
 
Dubinski, J.\ \& Carlberg, R.\ G.\ 1991, \apj, {\bf 378}, 496
 
Efstathiou, G., Lake, G.\ \& Negroponte, J.\ 1982, \mnras, {\bf 199}, 1069
 
Eisenstein, D.\ J.\ \& Loeb, A.\ 1995, \apj, {\bf 443}, 11
 
Elmegreen, D.\ M., Elmegreen, B.\ G.\ \& Bellin, A.\ D.\ 1990, \apj, {\bf
364}, 415.
 
Englmaier, P.\ \& Gerhard, O.\ E.\ 1998, in preparation
 
Evans, N.\ W.\ \& Read, J.\ C.\ A.\ 1998, \mnras, to appear
 
Fall, S.\ M.\ \& Efstathiou, G.\ 1980, \mnras, {\bf 193}, 189
 
Freeman, K.\ C.\ 1992, in {\it Physis of Nearby Galaxies, Nature or
Nurture?}, ed.\ T.\ X.\ Thuan et al.\ (Gif-sur-Yvette: Editions
Fronti\`eres), p~201
 
% Freudenreich, H.\ T.\ (1997) astro-ph/9707340
 
% Friedli, D., 1994, in {\it Mass-Transfer Induced Activity in Galaxies}, ed.
% I. Shlosman (Cambridge: Cambridge University Press) p~268
 
Friedli, D.\ \& Benz, W.\ 1993, \aap, {\bf 268}, 65
 
Gunn, J.\ E.\ 1982. In {\it Astrophysical Cosmology} p~233 eds H.\ A.\
Br\"uck, G.\ V.\ Coyne \& M.\ S.\ Longair (Vatican City: Pontificia Academia
Scientiarum)
 
Haehnelt, M.\ G.\ \& Rees, M.\ 1993, \mnras, {\bf 263}, 168
 
Hasan, H.\ \& Norman, C.\ 1990, \apj, {\bf 361}, 69
 
Helfer, T.\ T.\ \& Blitz, L., 1995, \apj, {\bf 450}, 90
 
Heller, C.\ H.\ \& Shlosman, I.\ 1994, \apj, {\bf 424}, 84
 
% Hernquist, L.\ \& Mihos, C.\ J.\ 1994, \apj, {\bf 431}, L9
 
Jablonka, J.\ \& Arimoto, N., 1994, \aap, {\bf 255}, 63
 
Kalnajs, A.\ J.\ 1983, in {\it Internal Kinematics and Dynamics of Galaxies},
IAU Symp.\ 100, ed E.\ Athanassoula (Dordrecht: Reidel) p~87
 
Kent, S.\ M.\ 1986, \aj, {\bf 91}, 1301
 
Knapen, J.\ H., Beckman, J.\ E., Heller, C.\ H., Shlosman, I. \& de Jong, R.\
S.\ 1995, \apj, {\bf 454}, 623
 
Kormendy, J.\ \& McClure, R.\ D.\ 1993, \aj, {\bf 105}, 1793
 
Kravtsov, A.\ V., Klypin, A.\ A., Bullock, J.\ S.\ \& Primack, J.\ R.\ 1998,
\apj, submitted -- astro-ph/9708176
 
Kuijken, K.\ 1995, in {\it Stellar Populations}, IAU Symposium {\bf 164}, eds
P.\ C.\ van der Kruit \& G.\ Gilmore (Dordrecht: Kluwer), 195
 
Lacey, C.\ G.\ \& Ostriker, J.\ P.\ 1985, \apj, {\bf 299}, 633
 
Lehnert, M.\ D., Heckman, T.\ M., Chambers, K.\ C.\ \& Miley, G.\ K.\ 1992,
\apj, {\bf 393}, 68
 
Lynden-Bell, D.\ 1969, Nature, {\bf 223}, 690
 
Magorrian, J.\ et al.\ 1998, \aj, {\bf 115}, 2285
 
Merritt, D.\ \& Quinlan, G.\ 1998, \apj, {\bf 498}, 625
 
Mestel, L.\ 1963, \mnras, {\bf 126}, 553.
 
Mo, H.\ J., Mao, S.\ \& White, S.\ D.\ M.\ 1998, \mnras, {\bf 295}, 319
 
Navarro, J.\ F.\ 1997, in {\it Dark and Visible Matter in Galaxies}, ed.\ M.\
Persic \& P.\ Salucci (San Francisco: ASP Conf.\ Ser.\ 117), 404
 
Navarro, J.\ F.\ \& Benz, M., 1991, \apj, {\bf 380}, 320
 
Navarro, J.\ F., Frenk, C.\ S.\ \& White, S.\ D.\ M.\ 1996, \apj, {\bf 462},
563
 
Navarro, J.\ F.\ \& Steinmetz, M., 1997, \apj, {\bf 478}, 13
 
Noguchi, M.\ 1987, \mnras, {\bf 228}, 635.
 
Noguchi, M.\ 1988, \aap, {\bf 203}, 259
 
Norman, C.\ A., Sellwood, J.\ A., \& Hasan, H.\ 1996, \apj, {\bf 462}, 114
 
Ostriker, J.\ P., \& Peebles, P.\ J.\ E.\ 1973, \apj, {\bf 186}, 467
 
Padovani, P., Burg, R.\ \& Edelson, R.\ A.\ 1990, \apj, {\bf 353}, 438
 
Palunas, P.\ \& Williams, T.\ B.\ 1998, \aj, submitted
 
% Papaloizou, J.\ C.\ B.\ \& Savonije, G.\ J., 1991, \mnras, {\bf 248}, 353
 
% Persic, M., Salucci, P.\ \& Stel, F.\ 1996, \mnras {\bf 281}, 27
 
Pfenniger, D.\ 1991, in {\it Dynamics of Disc Galaxies}, ed.\ B.\ Sundelius
(University of Gothenburg), p 191
 
Pfenniger, D., Combes, F.\ \& Martinet, L.\ 1994, \aap, {\bf 285}, 79
 
Prendergast, K.\ H.\ 1983, in {\it Internal Kinematics and Dynamics of
Galaxies}, IAU Symp.\ 100, ed E.\ Athanassoula (Dordrecht: Reidel) p~215
 
Quillen, A.\ C., Frogel, J.\ A., Kenney, J.\ D.\ P., Pogge, R.\ W.\ \& Depoy,
D.\ L.\ 1995, \apj, {\bf 441}, 549
 
Rees, M.\ J., 1997, astro-ph/9701161
 
Rubin, V.\ C., Ford, W.\ K., \& Thonnard, N.\ 1980, \apj, {\bf 238}, 471
 
Rubin, V.\ C., Kenney, J.\ D.\ P.\ \& Young, J.\ S., 1997, \aj, {\bf 113},
1250
 
Ryden, B.\ S.\ \& Gunn, J.\ E.\ 1987, \apj, {\bf 318}, 15
 
Sackett, P.\ D.\ 1997, \apj, {\bf 483}, 103
 
Schmidt, M., Schneider, D.\ P.\ \& Gunn, J.\ E.\ 1995, \aj, {\bf 110}, 68
 
Sellwood, J.\ A., 1981, \aap, {\bf 99}, 362
 
Sellwood, J.\ A.\ 1989, \mnras, {\bf 238}, 115
 
Sellwood, J.\ A.\ 1998, astro-ph/9711335
 
Sellwood J.\ A.\ \& Carlberg R.\ G.\ 1984, \apj, {\bf 282}, 61
 
Sellwood J.\ A.\ \& Valluri, M.\ 1997, \mnras, {\bf 287}, 124
 
Sellwood, J.\ A.\ \& Wilkinson, A.\ 1993, Rep. Prog. Phys., {\bf 56}, 173
 
Shaver, P.\ A., Wall, J.\ V., Kellerman, K.\ I., Jackson, C.\ A.\ \& Hawkins,
M.\ R.\ S.\ 1996, Nature, {\bf 384}, 439
 
Shlosman, I., Begelman, M.\ C.\ \& Frank, J., 1990, Nature, {\bf 345}, 679
 
% Shlosman, I., Frank, J.\ \& Begelman, M.\ C., 1989, Nature, {\bf 338}, 45
 
Silk, J.\ \& Rees, M.\ J., 1998, astro-ph/9801013
 
Small, T.\ A.\ \& Blandford, R.\ D.\ 1992, \mnras, {\bf 259}, 725
 
% Syer, D., Mao, S.\ \& Mo, H.\ J.\ 1997, astro-ph/9711160
 
Taylor, G.\ L., Dunlop, J.\ S., Hughes, D.\ H.\ \& Robson, E.\ I.\ 1996,
\mnras, {\bf 283}, 930
 
Toomre, A.\ 1981, in {\it Structure and Evolution of Normal Galaxies}, eds
S.\ M.\ Fall \& D.\ Lynden-Bell (Cambridge: Cambridge University Press), 111
 
Toomre, A.\ 1990, in {\it Dynamics \& Interactions of Galaxies}, ed.\ R.\
Wielen, (Berlin, Heidelberg: Springer-Verlag), 292
 
van Albada, T.\ S.\ \& Sancisi, R.\ 1986, Phil Trans London A, {\bf 320}, 447
 
van der Kruit, P.\ C.\ 1987, \aap, {\bf 173}, 59
 
van der Kruit, P.\ C.\ 1995, in {\it Stellar Populations}, IAU Symposium {\bf
164}, eds P.\ C.\ van der Kruit \& G.\ Gilmore (Dordrecht: Kluwer), 205
 
Villumsen, J.\ V.\ \& Gunn, J.\ E.\ 1987, unpublished
 
Weinberg, M.\ D.\ 1985, \mnras, {\bf 213}, 451
 
Weiner, B., Sellwood, J.\ A., Williams, T.\ B.\ \& van Gorkom, J.\ 1998, in
preparation
 
White, S.\ D.\ M.\ \& Rees, M.\ J.\ 1978, \mnras, {\bf 183}, 341
 
Worthey, G.\ 1994, \apj Supp, {\bf 95}, 107
 
Zang, T.\ A.\ 1976, PhD Thesis, MIT
 
% Zwaan, M.\ A., van der Hulst, J.\ M., de Blok, W.\ J.\ G.\ \& McGaugh, S.\
% S.\ 1995, \mnras, {\bf 273}, L35
 
\end{references}
\end{document}